\documentclass[prb,twocolumn,showpacs,preprintnumbers,superscriptaddress,unsortedaddress,runinaddress,amsmath,amssymb]{revtex4}

\usepackage{graphicx}
\usepackage{dcolumn}
\usepackage{bm}

\begin{document}


\title{Electronic transport in ferromagnetic alloys and the Slater-Pauling Curve}

\author{S. Lowitzer}
\author{D. K\"odderitzsch}
\author{H. Ebert}
\affiliation{%
Department Chemie und Biochemie, Physikalische Chemie, Universit\"at M\"unchen, Butenandstr. 5-13, 81377 M\"unchen, Germany\\
}%

\author{J. B. Staunton}
\affiliation{
Department of Physics, University of Warwick, Coventry CV4 7AL, United Kingdom\\
}%

\date{\today}

\begin{abstract}
Experimental measurements of the residual resistivity $\rho(x)$ of the binary alloy system Fe$_{1-x}$Cr$_x$ 
have shown an anomalous concentration dependence which deviates
significantly from Nordheim's rule. In the low ($x
< 10\%$) Cr concentration regime the resistivity has been found to increase linearly
with $x$ until  $\approx$ 10\% Cr where the resistivity
reaches a plateau persisting to $\approx$ 20\% Cr.  In this paper we present
$ab$-$initio$ calculations of $\rho(x)$ which explain this anomalous behavior and which are
based on the Korringa-Kohn-Rostoker (KKR) method in conjunction with the
Kubo-Greenwood formalism. Furthermore we are able to show that the effects of short-range
ordering or clustering have little effect via our use of the nonlocal
coherent-potential approximation (NL-CPA). For the
interpretation of the results we study the alloys' electronic structure by
calculating the Bloch spectral function particularly in the vicinity of the Fermi 
energy. From the analysis of our results we infer that a 
similar behavior of the resistivity should also be obtained for iron-rich Fe$_{1-x}$V$_x$
alloys - an inference confirmed by further explicit resistivity calculations. 
Both of these alloy systems belong to the same branch of the famous Slater-Pauling plot and we
postulate that other alloy systems from this branch should show a similar behavior. 
Our calculations show that the
appearance of the plateau in the resistivity can be attributed to the dominant
contribution of minority spin electrons to the conductivity which is nearly unaffected by
increase in  Cr/V concentration $x$ and we remark that this minority
spin electron feature is also responsible for the simple linear 
variation of the average moment in the Slater-Pauling plot for these
materials.

\end{abstract}

\maketitle

\section{Introduction}
The Slater-Pauling plot of average magnetisation per atom $M$ versus
valence electron number $N_v$ plays a pivotal role in the
understanding of the properties of ferromagnetic
alloys.~\cite{Boz51} Its triangular structure of two straight
lines with gradients of opposite sign neatly categorises most alloys
into one of two classes where $\frac{dM}{dN_v} \pm 1$. Long ago
Mott~\cite{Mot64} pointed out how this behavior can be explained by
requiring either the number of majority or minority spin electrons to
be fixed. This notion has subsequently been confirmed and given
substance by modern spin density functional theory (DFT)
calculations~\cite{Kub00,Sta94,MWM84,JPS87}. In this paper we
follow the implications of this for the residual resistivities of
ferromagnetic alloys. In particular, using the latest $ab$-$initio$
techniques for describing disordered systems, we demonstrate how the
measured apparently anomalous resistivities of iron-rich Fe$_{1-x}$Cr$_x$
alloys~\cite{Nikolaev99,Tsiovkin05} are directly attributable to their
average minority-spin electron number being fixed and consequently
their location on the $\frac{dM}{dN_v}= +1$ section of the
Slater-Pauling plot. We postulate that other alloys in this category
should also show the same behavior and we strengthen this conjecture
by the findings from another detailed $ab$-$initio$ study of
Fe$_{1-x}$V$_x$ alloys. We also infer that short-range order should
have little effect on the resistivities of alloys in this category and
again we are able to back up these remarks by detailed specific
$ab$-$initio$ calculations.

Many DFT calculations for disordered ferromagnetic alloys show that
the majority-spin electrons `see' little disorder and that the
majority-spin $d$-states are fully occupied. This leads to
$\frac{dM}{dN_v}= -1$ ~\cite{Kub00}. In contrast the minority-spin
electron states are significantly affected by disorder. The overall
electronic transport is therefore taken up by the $sp$-majority-spin
electrons. Alloys in this category include fcc-based CoMn, FePt
and Ni-rich NiFe alloys.

\begin{figure} 
\includegraphics[scale=0.4,clip]{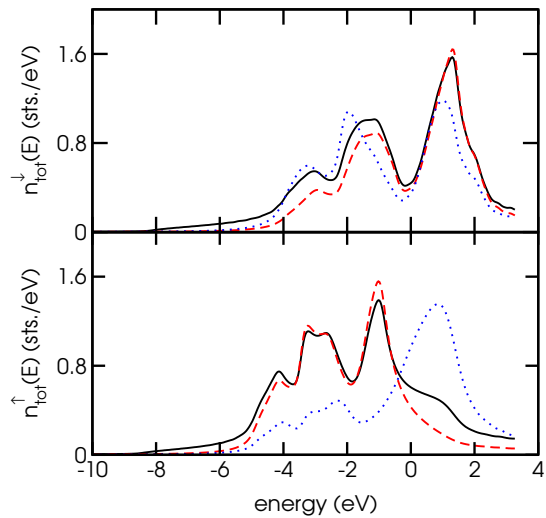}
 \caption{\label{plot:DOS1}(Color online) Spin projected CPA density of states of disordered bcc Fe$_{0.8}$Cr$_{0.2}$. The solid (black) line shows the total DOS, the dashed (red) line shows the Fe $d$-states and the dotted (blue) line shows the Cr $d$-states. The Fermi level is located at the zero of the energy-axis.}
\end{figure}

On the other hand, for some other alloys, typically Fe-rich,
bcc-based alloys and many Heusler alloys, the number of
minority-spin $d$-electrons is fixed as the Fermi energy $E_F$ is pinned
at a low level in a trough of the $d$-electron density of states. $\frac{dM}{dN_v}= +1$
of the Slater-Pauling curve follows directly from
this~\cite{Kub00}. It is the ramifications of this feature for the
electronic transport in such alloys that we investigate here. This
time disorder is `seen' strongly by the majority-spin electrons and
rather weakly by the minority-spin electrons. Fig. \ref{plot:DOS1}
provides a relevant illustration for a bcc  Fe$_{0.8}$Cr$_{0.2}$
disordered alloy.~\cite{Sta94} The Fe- and Cr-related minority-spin
densities of states have similar structure in contrast to those of
majority spin. $E_F$ is positioned in a valley making the average
number of minority spin electrons $\approx$ 3. Moreover from these
observations we can expect the resistivity to be dominated by minority
spin electrons and to be rather insensitive to overall composition and
short-range order. 
Recently, measurements of the residual electrical resistivity of
iron-rich Fe$_{1-x}$Cr$_x$ alloys have been reported and described as
anomalous. The measurements show that the resistivity increases as
small amounts of Cr are added to Fe until a plateau is reached ranging
from $x = 10\%$ to $20\%$~\cite{Tsiovkin05}. This behavior differs
markedly from the Nordheim parabolic concentration dependence. An
analysis of the data is hindered by a complexity of short-range order
in Fe$_{1-x}$Cr$_x$. \citet{Mirebeau84} reported that for $x < 10\%$
the system develops short-ranged order whereas for larger $x$
short-ranged clustering is found. At higher Cr concentrations still ($
> 20\%$), the alloys can undergo on aging a separation into Fe-rich
($\alpha$) and Cr-rich ($\alpha^{\prime}$)
phases~\cite{Cieslak00,Olsson06} leading either to a miscibility gap
or a transformation into a tetragonal $\sigma$ phase.

We use the Kubo-Greenwood formalism~\cite{Kub57,Gre58}
implemented with the Korringa-Kohn-Rostoker (KKR) method for $ab$-$initio$
calculations of the residual resistivities of Fe$_{1-x}$Cr$_x$
alloys. Disorder is accounted for by the use of the coherent potential
approximation (CPA)~\cite{But85}. In order to examine the effects of
short-ranged order within the disorder we use our recently developed
method which involves the non-local CPA (NL-CPA)~\cite{TSE+07}. We
complement our $ab$-$initio$ resistivity calculations with Bloch spectral
function calculations which show the electronic structure in the
vicinity of $E_F$ and enable a connection to be made to the
semi-classical Boltzmann description of electronic transport
properties~\cite{BS82}.

We find that our calculations fully support our initial expectations
outlined above which are linked to the alloys' location on the
Slater-Pauling plot in addition to describing the experimental data
well. Moreover from further calculations for Fe$_{1-x}$V$_x$ alloys we
predict that a similar behavior will be found for these materials as
well as for other systems sited in this region of the Slater-Pauling
plot.

The next section outlines the $ab$-$initio$ theoretical framework
(full details are given in Refs. \onlinecite{But85} and
\onlinecite{TSE+07}). This is followed by our study of the resistivity
of Fe$_{1-x}$Cr$_x$ as a function of $x$ together with Bloch spectral
function interpretation. We then describe a similar investigation of
Fe$_{1-x}$V$_x$ alloys. The last section summarises and emphasises the
overall conclusions.
\section{Theory}
\subsection{Kubo-Greenwood equation}
The basis of our investigations is supplied by the Kubo-Greenwood
equation for the symmetric part of the conductivity tensor
\cite{But85}
\begin{equation}
\label{eq:KG}
\sigma_{\mu\nu}= \frac{\hbar}{\pi N\Omega} \, {\rm{Tr}}\,\bigg\langle \hat{j}_{\mu} \, \Im G^+(E_F) \, \hat{j}_{\nu} \, \Im G^+(E_F) \bigg\rangle.
\end{equation}
Here $N$ is the number of atomic sites, $\Omega$ the volume per atom,
$\langle\rangle$ indicates an averaging over configurations and $\Im
G^+(E_F)$ is the imaginary part of the retarded one particle
Green function at the Fermi energy $E_F$. For the determination
of $\Im G^+(E_F)$ we use multiple scattering theory (MST) which
is the basis of the KKR band structure method. Within MST the real
space representation of $\Im G^+$ has the
following form \cite{FS80}
\begin{equation}
\Im G^+(\vec{r},\vec r\,',E)= 
 \Im \sum_{\Lambda_1 \Lambda_2}
Z_{\Lambda_1}(\vec r,E)
 \tau_{\Lambda_1 \Lambda_2}(E)
Z_{\Lambda_2}^\times(\vec r\,',E)\,.
\end{equation}
Using a fully relativistic implementation, the wave functions
$Z(Z^\times)$ are the regular right (left) hand side solutions of the
Dirac equation, $\tau$ is the scattering path operator and $\Lambda =
(\kappa,\mu)$ with $\kappa$ and $\mu$ being the relativistic
spin-orbit and magnetic quantum numbers \cite{Ros61}. Finally,
$\hat{j}_{\mu}$ in Eq. (\ref{eq:KG}) is the current density operator
that is given in relativistic form by 
\begin{equation}
\label{eq:operator}
\hat{j}_{\mu} = ec\,{\bm \alpha}_{\mu}
\end{equation}
where ${\bm \alpha}_{\mu}$ is one of the standard Dirac matrices
\cite{Ros61}. \\ The central step to calculate the conductivity is to
handle the averaging over all possible configurations. For a randomly
disordered system the CPA \cite{Sov67} is a well established method
for the configuration averaging. The CPA introduces an effective
medium which represents the electronic properties of the disordered
alloy. This effective medium is described by its single site $t$
matrix $t_{{\rm CPA}}$ in combination with its averaged scattering
path operator $\tau_{{\rm CPA}}$
\begin{equation}
\underline{\tau}_{{\rm CPA}}  =  \frac{1}{\Omega_{{\rm BZ}}}\int_{\Omega_{{\rm BZ}}}d^{3}k\left[ (\underline{t}_{{\rm CPA}})^{-1}-\underline{G}(\vec{k},E)  \right]^{-1} 
\end{equation}
where the underline indicates matrices with respect to the
spin-angular character $\Lambda$.  Assuming a binary alloy
A$_{1-x}$B$_x$, $\underline{t}_{{\rm CPA}}$ and $\underline{\tau}_{{\rm
CPA}}$ have to fulfill the CPA condition
\begin{equation}
\label{eq:CPA-cond}
\underline{\tau}_{{\rm CPA}} = (1-x)\,\underline{\tau}_{{\rm A}} + x\,\underline{\tau}_{{\rm B}}\,.
\end{equation}
$\underline{\tau}_{{\rm A(B)}}$ is the scattering path operator for an
atom of type A(B) embedded into the CPA medium and
$\underline{G}(\vec{k},E)$ is the KKR structure constants matrix. With
these CPA equations it is possible to construct the effective medium
iteratively.\\ \citet{But85} derived within the CPA a scheme to handle
the averaging over configurations of two Green functions. His
formulation of the Kubo-Greenwood equation on the basis of the KKR-CPA
is widely applied and gives in general good agreement with
experimental data \cite{BE95,BE95a,EVB96,Banhart96,DBC00,VEB03}. \\ It
has to be emphasized that the CPA is a single site theory and
therefore neglects correlations concerning the occupation of the
neighboring atomic sites. For that reason one has to use more
elaborate methods to include SRO effects. One possibility is to use
the NL-CPA \cite{JK01,RSG03,BGJ+05,REG+06,Row06}. The NL-CPA creates
(as the CPA) an effective medium and can be understood as a cluster
generalization of the well established CPA. The corresponding NL-CPA
equations can be written as:
\begin{eqnarray}
\underline{\tau}^{IJ}_{{\rm NLCPA}} & = & \frac{1}{\Omega_{{\rm BZ}}}\sum_{\vec{K}_n}\left(\int_{\Omega_{\vec{K}_n}}d^{3}k\left[ (\underline{t}_{{\rm NL-CPA}})^{-1}  - \underline{G}(\vec{k},E) \right. \right. \nonumber \\ 
& & \left. \left.-\delta\underline{\hat{G}}(\vec{K}_n,E)  \right]^{-1}\right) e^{i\vec{K}_n\cdot(\vec{R}_I-\vec{R}_J)}\\
\label{eq:NL-CPA-cond}
\underline{\underline{\tau}}_{{\rm NLCPA}} & = & \sum_{\gamma}P_{\gamma}\underline{\underline{\tau}}_{\gamma} \;\;\;\;\;\;\;\;\;({\rm with}\;  \sum_{\gamma}P_{\gamma} = 1)\,,
\end{eqnarray}
with the double underline indicating matrices with respect to site and
spin-angular character indices $I$ and $\Lambda$, respectively. In
Eq. (\ref{eq:NL-CPA-cond}) $\underline{\underline{\tau}}_{\gamma}$ is
the scattering path operator for a cluster of type $\gamma$ embedded
into the NL-CPA medium, $P_{\gamma}$ is the probability that this
cluster type occurs and $\delta\underline{\hat{G}}(\vec{K}_n,E)$ are
effective structure constant corrections for tile $\vec{K}_n$ which
account for nonlocal correlations due to the disorder configurations
\cite{RSG03}. \\ Recently the KKR-NL-CPA was combined with the
Kubo-Greenwood equation \cite{TSE+07}. On the basis of this scheme it
is possible to investigate ordering effects on the residual
resistivity in a systematic way.\\ For ferromagnetic cubic solids with
the magnetization along the $z$-axis the conductivity tensor consists
of five independent components:
$\sigma_{xx}=\sigma_{yy}=\sigma_{\perp}$,
$\sigma_{zz}=\sigma_{\parallel}$ and
$\sigma_{xy}=-\sigma_{yx}$. $\sigma_{\perp}$ and $\sigma_{\parallel}$
are the transverse and longitudinal conductivities, while
$\sigma_{xy(yx)}$ determine the spontaneous or anomalous Hall
resistivity. In line with the above mentioned experimental
investigations only the isotropic resistivity $\overline{\rho}$ is
considered that is given by
\begin{equation}
\overline{\rho}=\frac{1}{3}(2\rho_{\perp}+\rho_{\parallel})
\end{equation} 
with $\rho_{\perp}=\sigma_{\perp}^{-1}$ and
$\rho_{\parallel}=\sigma_{\parallel}^{-1}$ if the spin-orbit induced
component $\sigma_{xy}$ is ignored.
\subsection{Bloch spectral functions}
In a most general way the density of states may be defined as
\cite{Eco06}
\begin{equation}
n(E)= \sum_n\delta(E-E_n),
\end{equation}
where $E_n$ are the electronic eigenvalues of the system. In analogy
 the Bloch spectral function (BSF) can be defined by \cite{FS80}:
\begin{equation}
A(E,\vec{k})= \sum_n\delta[E-E_n(\vec{k})]
\end{equation}
and for that reason can be regarded as a $\vec{k}$-resolved density of
state. Dealing with an ordered system and a given $\vec{k}$-vector the
BSF has at the positions of the eigenvalues an infinitely sharp peak
and everywhere else it is zero. If one has an alloy instead of a
perfect crystal an appropriate expression for the BSF within KKR-CPA
was worked out by \citet{FS80}.
\begin{eqnarray}
\label{eq:BSF}
A(E,\vec{k}) & = & -\frac{1}{\pi}\Im\,{\rm Tr}\left[\underline{F}^c\underline{\tau}_{{\rm CPA}}(E,\vec{k})\right] \nonumber \\ 
& & -\frac{1}{\pi}\Im\,{\rm Tr}\left[(\underline{F}^c-\underline{F}^{cc})\underline{\tau}_{{\rm CPA}}\right]
\end{eqnarray}
with
\begin{equation}   
\underline{\tau}_{{\rm CPA}} =  \frac{1}{\Omega_{{\rm BZ}}}\int_{{\rm BZ}}d^3k \,\underline{\tau}_{{\rm CPA}}(E,\vec{k})\,. 
\end{equation}
The matrices $\underline{F}^c$ and $\underline{F}^{cc}$ are given in
terms of the overlap integrals 
\begin{equation}
F^{\alpha\beta}_{\Lambda \Lambda^{\prime}}  =  \int_{\Omega} d^3r\,Z_{\Lambda}^{\alpha\times}(E,\vec{r})\,Z_{\Lambda^{\prime}}^{\beta}(E,\vec{r})\,. 
\end{equation}
$\alpha,\beta$ denotes an atom type of the alloy. For more details
and explicit expressions see Ref. \onlinecite{FS80}. Compared to a pure
system, the BSF for an alloy becomes broadened due to the disorder.
This broadening can be related to the lifetime of an electron in a
Bloch state and is therefore quite useful for the interpretation of
resistivity data \cite{BS82}. \\ With the BSF it is possible to
discuss a dispersion relation $E(\vec{k})$ even for alloys
\cite{Fau82}. Strictly spoken such a dispersion relation is in general
not defined for alloys because $\vec{k}$ is not a good quantum number
for disordered systems.  Nevertheless, the dispersion relation
represented by the BSF can be used to calculate Fermi velocities
\cite{BS82} and gives therefore useful hints for the interpretation of
resistivity data. 
\subsection{Computational details}
All calculations were done in the framework of spin density functional
theory using the local spin density approximation (LSDA) with the
parameterization of \citet{VWN80} for the exchange correlation
potential. \\ The CPA and the NL-CPA formulations of the
Kubo-Greenwood equation were implemented within the self-consistent,
spin polarized, relativistic KKR (SPR-KKR) scheme \cite{Ebe00}. All
calculations include vertex corrections \cite{TSE+07,But85}. To ensure
convergence of the results with respect to the angular momentum
expansion the cutoff $l_{max}=4$ has been used.
\section{Results}
\subsection{CPA Results}
\begin{figure} \begin{center}
 \includegraphics[scale=0.36]{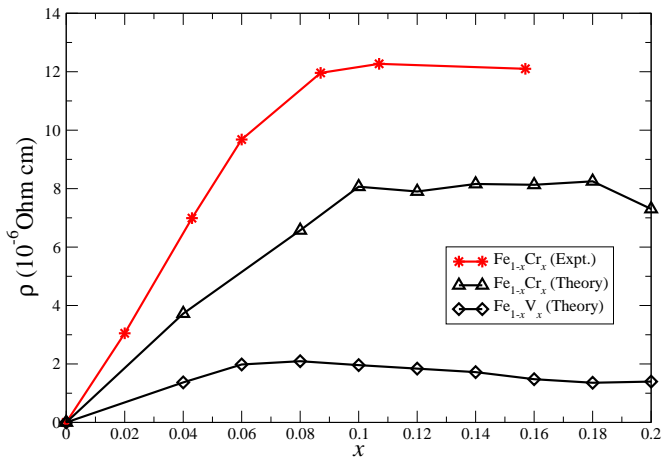}
 \caption{\label{plot:FeCr_res}(Color online) The residual resistivity of Fe$_{1-x}$Cr$_x$ and Fe$_{1-x}$V$_x$ as a function of the Cr/V concentration $x$. The asterisks (red line) show the experimental data for Fe$_{1-x}$Cr$_x$ of Ref. \onlinecite{Nikolaev99} at 4.2K. The triangles and diamonds (black lines) show our CPA results for Fe$_{1-x}$Cr$_x$ and Fe$_{1-x}$V$_x$, respectively.}
 \end{center}
\end{figure}

\begin{figure}
 \begin{center}
 \includegraphics[scale=0.4,angle=270,clip]{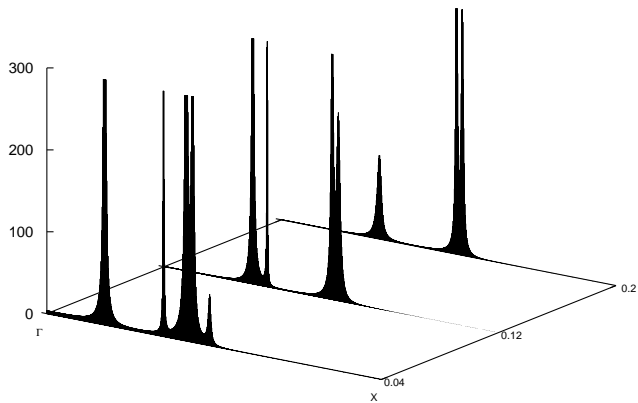}
 \caption{\label{plot:BSF_peaks}BSF along the $\Gamma$ - X direction for the minority spin subsystem in Fe$_{1-x}$Cr$_x$ for three different Cr concentrations (4\%, 12\% and 20\% Cr). The cusps of the BSF have been cut at 300 a.u..}
 \end{center}
\end{figure}

\begin{figure}
 \begin{center}
\includegraphics[scale=0.53,angle=270,clip]{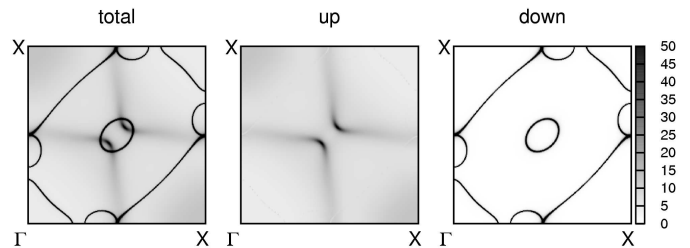}
 \caption{\label{plot:BSF_FeV}Total and spin projected BSF of Fe$_{0.8}$V$_{0.2}$ at the Fermi energy in the (001) plane. The black regions correspond to values $>$ 50 a.u.. For a better resolution the cusps of the BSF have been cut.}
 \end{center}
\end{figure}

\begin{figure}
 \begin{center}
 \includegraphics[scale=0.4,angle=270,clip]{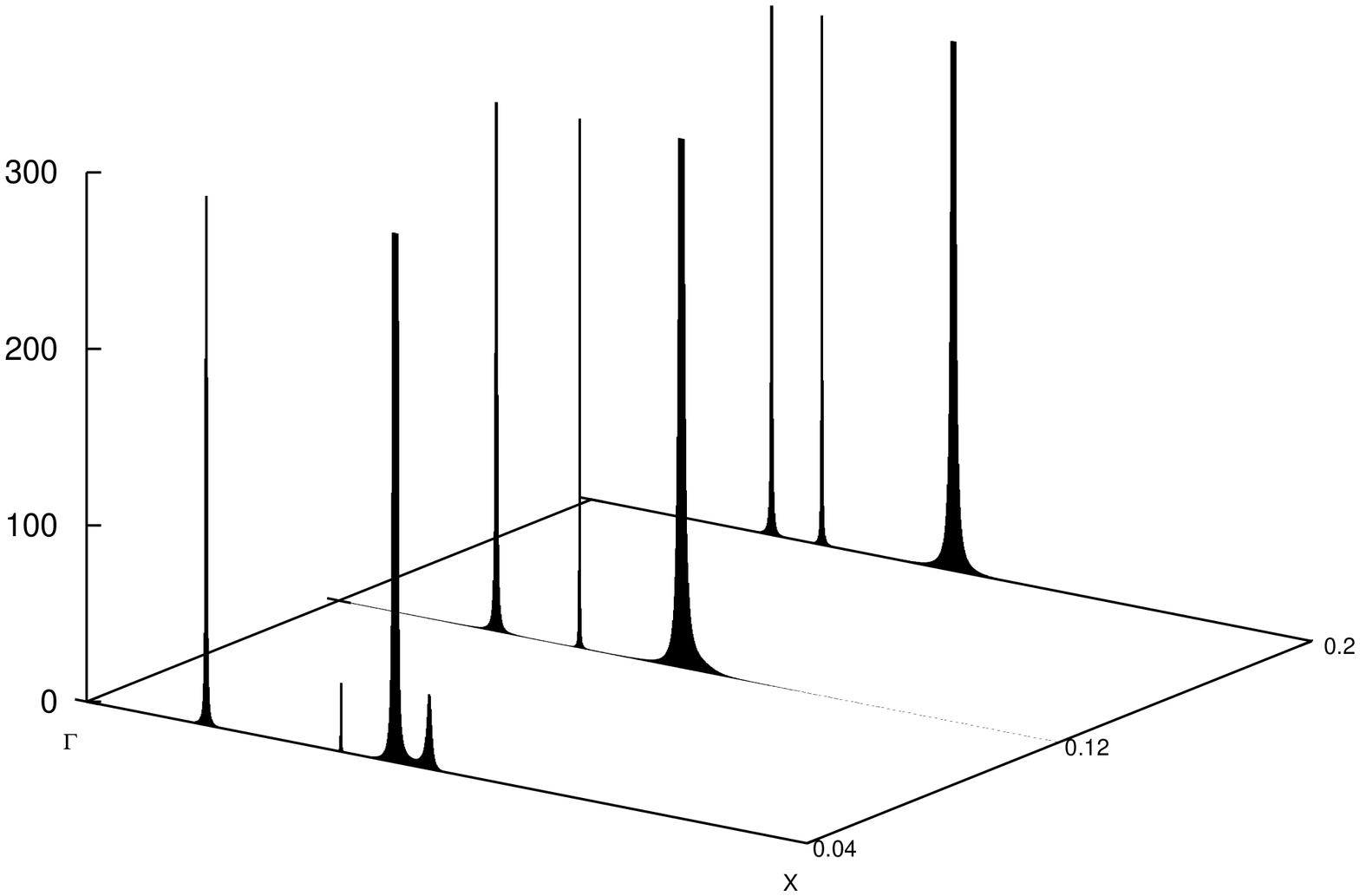}
 \caption{\label{plot:BSF_peaks_FeV}BSF along the $\Gamma$ - X direction for the minority spin subsystem in Fe$_{1-x}$V$_x$ for three different V concentrations (4\%, 12\% and 20\% V). The cusps of the BSF have been cut at 300 a.u..}
 \end{center}
\end{figure}

The central result of our calculations is shown in
Fig. \ref{plot:FeCr_res}.
This figure shows the residual resistivity of Fe$_{1-x}$Cr$_x$ and
Fe$_{1-x}$V$_x$ as a function of the Cr/V concentration $x$. As
mentioned above, the experimental Fe$_{1-x}$Cr$_x$ data show an
anomalous behavior. In the low Cr concentration regime ($< \approx
10\%$) the residual resistivity increases with increasing Cr
concentration. Further increase of the Cr content does not lead to a
further increase of the resistivity. Our theoretical results show the
same variation with the Cr concentration. At 10\% Cr we obtain the
highest value for the resistivity. If one further increases the Cr
concentration, the resistivity stays more or less constant at $\approx
8 \mu \Omega cm$. \\ The Fe$_{1-x}$V$_x$ alloys show a similar
behavior for the theoretical residual resistivity. With increasing V
concentration the residual resistivity increases up to $\approx$
2$\mu\Omega$cm (at 6\% V). Further increase of the V concentration
leads only to small changes in the residual resistivity. \\ This
behavior can be explained by a different variation of the electronic
structure for the majority and minority spin subsystems when the Cr/V
concentration changes. Adding Cr to pure Fe in a random way the
disorder in the system increases and with this the resistivity
increases. This conventional behavior is observed in the regime with a
Cr content smaller than $\approx 8\%$ Cr where the system shows a
Nordheim like behavior. To identify the contribution of the
majority/minority spin subsystems to the conductivity, we calculated
the BSF according to Eq. (\ref{eq:BSF}).
\begin{figure}
 \begin{center}
 \includegraphics[scale=0.53,angle=270,clip]{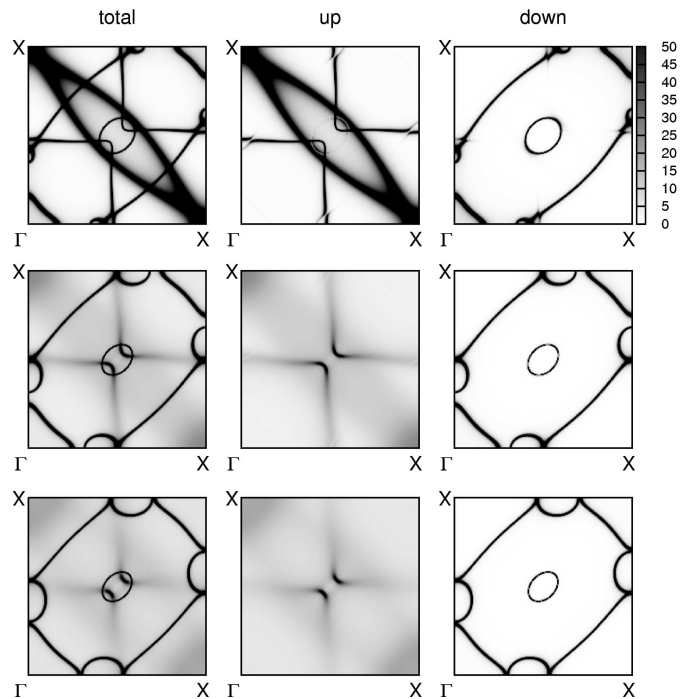}
 \caption{\label{plot:BSF}Total and spin projected BSF of Fe$_{1-x}$Cr$_x$ at the Fermi energy in the (001) plane for different Cr concentrations (top: 4\% Cr, middle: 12\% Cr, bottom: 20\% Cr). The black regions correspond to values $>$ 50 a.u.. For a better resolution the cusps of the BSF have been cut.}
 \end{center}
\end{figure}
Fig. \ref{plot:BSF} shows the total and spin projected BSF for three
different Cr concentrations (4\%, 12\% and 20\% Cr). The important
observation from the displayed BSF are the different dependencies of
the majority and minority spin subsystem on the Cr concentration. At
4\% Cr both spin subsystems show sharp peaks for the BSF which
indicate that the impact of disorder is relatively small. If one
increases the Cr concentration up to 12\% a dramatic change
occurs. For the BSF of the majority subsystem the prominent
lens-shaped band disappears and the remaining rectangular-shaped band
become strongly smeared out whereas the minority component is almost
unchanged. Further increase of the Cr concentration continues this
trend. For a better illustration of the influence of the Cr increase
on the minority component we show in Fig. \ref{plot:BSF_peaks}
explicitly the peaks of the BSF at the Fermi energy along the
$\Gamma-X$ direction.
\begin{figure}
 \begin{center}
\includegraphics[scale=0.36,clip]{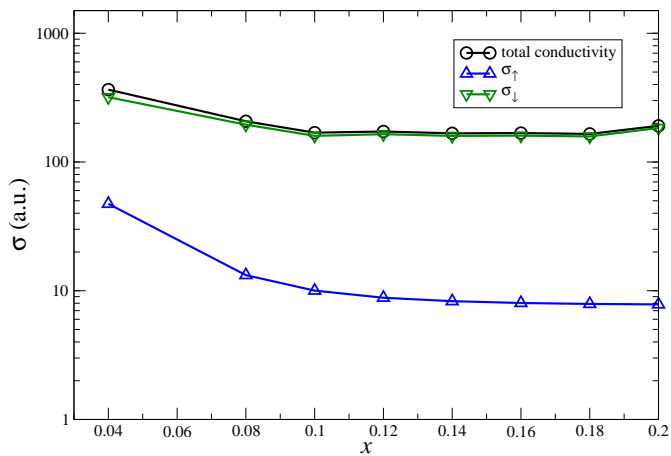}
 \caption{\label{plot:FeCr_spin}(Color online) Spin resolved conductivities $\sigma_{\uparrow}$ and $\sigma_{\downarrow}$ for Fe$_{1-x}$Cr$_x$}
 \end{center}
\end{figure}
The BSF shows three main peaks with the last peak (the closest peak to
the $X$ point) being split. At 4 \% Cr an additional small peak is
present to the right of the split peak. This peak is due to a
hybridisation with the majority subsystem. This can be inferred from
Fig. \ref{plot:BSF} if one compares the minority and majority BSF
for 4 \% Cr. The reason for this hybridisation is, that in fully
relativistic calculations the spin is not a good quantum number
because of the presence of spin-orbit coupling \cite{EVB97a}. \\ The
behavior of the three remaining peaks is quite different. The 
narrow peak shifts with increasing Cr concentration towards the
$\Gamma$-point and at 20\% Cr this peak overlaps with the peak closest
to the $\Gamma$-point. The split peak remains nearly fixed at its
position but one observes a narrowing with increasing Cr
concentration, which corresponds to an increased lifetime of this
state. \\ For comparison we also calculated BSF for Fe$_{1-x}$V$_x$
and found a similar behavior of the majority/minority spin subsystem
as for Fe$_{1-x}$Cr$_x$. Fig. \ref{plot:BSF_FeV} shows the spin
projected BSF for Fe$_{0.8}$V$_{0.2}$. One can see that again the
majority component becomes smeared out whereas the minority component
displays sharp peaks. To get a more detailed picture of the
Fe$_{1-x}$V$_x$ BSF we show in Fig. \ref{plot:BSF_peaks_FeV} a similar
picture as shown in Fig. \ref{plot:BSF_peaks} for Fe$_{1-x}$Cr$_x$. If
one compares Fig. \ref{plot:BSF_peaks_FeV} with
Fig. \ref{plot:BSF_peaks} one can see that for Fe$_{1-x}$V$_x$ the BSF
peaks are more sharp than for Fe$_{1-x}$Cr$_x$.  Therefore
one can say that the minority spin electrons ``see'' a smaller
difference between Fe and V atoms compared to Fe and Cr atoms. This
explains why the residual resistivities are higher in Fe$_{1-x}$Cr$_x$
compared to Fe$_{1-x}$V$_x$. Fig. \ref{plot:BSF_peaks_FeV} shows that
the increased disorder due to the increased V concentration do not
affect the BSF peaks of the minority spin subsystem. \\ To identify
the character of the smeared out states from the majority subsystem,
we projected the Fe$_{1-x}$Cr$_x$ BSF according to its $s$-, $p$-
and $d$-contributions.
\begin{figure}
 \begin{center}
 \includegraphics[scale=0.53,angle=270,clip]{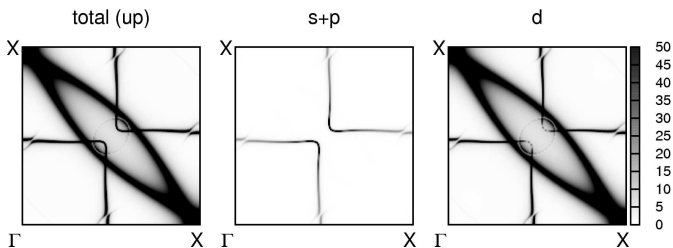}
 \caption{\label{plot:BSF_sp_d}Projected majority component of the Fe$_{1-x}$Cr$_x$ BSF at 4\% Cr. The left plot shows the total BSF whereas the middle and the right plot show the $s+p$ and $d$ projected BSF, respectively.}
 \end{center}
\end{figure}
This is shown in Fig. \ref{plot:BSF_sp_d}. The main part of the
majority states has  $d$-like character. These states obviously
strongly broaden with increasing Cr concentration. This behavior is
opposite to that of the minority subsystem although this is also
dominated by $d$-like states. The different behavior of the $d$-states
for the two spin subsystems can also be seen in the density of states.
\begin{figure}
 \includegraphics[scale=0.4,clip]{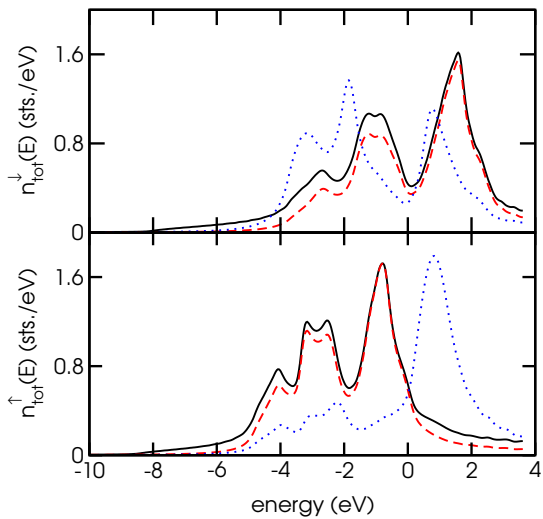}
 \caption{\label{plot:DOS}(Color online) Spin projected CPA density of states of Fe$_{0.96}$Cr$_{0.04}$. The solid (black) line shows the total DOS, the dashed (red) line shows the Fe d-states and the dotted (blue) line shows the Cr d-states. The Fermi level is located at the zero of the energy-axis.}
\end{figure}
In Figs. \ref{plot:DOS1} and \ref{plot:DOS} the spin projected DOS of
Fe$_{1-x}$Cr$_x$ for two different Cr concentrations is shown. In
addition to the total DOS we show the $d$-like part of the DOS. The
DOS shown in Fig. \ref{plot:DOS} is very close to that of pure Fe. One
can see that for the majority component also the antibonding Fe
$d$-states are occupied whereas for the minority component the Fermi
level is located in a so-called pseudogap below the antibonding
states. Fig. \ref{plot:DOS1} clearly shows the relative positions of
the Cr and Fe $d$-states. These states are strongly hybridized for the
minority component. The opposite happens in the majority spin channel,
where the Fe and the Cr $d$-states are well separated in energy. \\
With increasing Cr concentration the antibonding Fe $d$-peak of the
majority component becomes more and more depopulated and new states
appear above the Fermi level. \citet{Olsson06} showed that this leads
to a completely smeared out band at approximately equiatomic
composition. If one compares this with the behavior of the $d$-states
of the minority component, one can see that the increase in Cr
concentration has no effect on the total DOS of the minority
component.  \\ It is well known that the DOS of Fe$_{1-x}$V$_x$
consists of a minority spin subsystem where the Fermi energy is pinned
in a pseudogap and a majority spin subsystem which becomes broadened
and depopulated with increasing V concentration
\cite{Sta94,JPS87}. These similar characteristics of the
Fe$_{1-x}$Cr$_x$ and the Fe$_{1-x}$V$_x$ DOS are responsible for the
appearance of the Slater-Pauling curve for the magnetic moment of
these alloys. Due to the fact that the number of minority electrons
($N_{\downarrow}$) is independent of the Cr/V concentration the
magnetization per atom $M$ varies linearly with the Cr/V concentration
\cite{Sta94}
\begin{equation}
M = Z - 2N_{d\,\downarrow} - 2N_{sp\,\downarrow}\,,
\end{equation}
with $Z$ the number of valence electrons. The number of $sp$-electrons
in the minority spin system $N_{sp\,\downarrow}$ changes only very
little across the 3$d$ row \cite{Sta94}). Therefore one can conclude
that the Cr/V concentration independent hybridized Fe and Cr/V
$d$-states of the minority spin subsystem are responsible for the
appearance of the Slater-Pauling curve and in addition for the
apparently anomalous behavior of the residual resistivity of these
materials. \\
The discussion of the electronic structure of Fe$_{1-x}$Cr$_x$ and
Fe$_{1-x}$V$_x$ in terms of the BSF and DOS curves has obviously been
made in the spirit of the two-current model for the conductivity of
spin-polarized solids. The fully relativistic approach for calculating
the conductivity used here strictly spoken doesn't allow a
decomposition of the conductivity into spin channels. The reason for
this is that spin-orbit coupling not only gives rise to the
off-diagonal elements $\sigma_{xy}$ mentioned above but also to
spin-flip contributions that influence the isotropic conductivity or
resistivity, respectively. If the spin-orbit coupling strength is not
too strong, however, an approximate decomposition can nevertheless be
made. As was demonstrated in Ref. \onlinecite{PEP+04}, ignoring the
small spin-flip elements connected with the current density operator
in Eq. (\ref{eq:operator}), one can still define a spin-projected
conductivity. This is indeed justified for Fe$_{1-x}$Cr$_x$ by the
fact that the sum of the approximate conductivities $\sigma_{\uparrow}
+ \sigma_{\downarrow}$ hardly differs from the conductivity calculated
in a relativistic way (Fig. \ref{plot:FeCr_res}).  The results for
$\sigma_{\uparrow}$ and $\sigma_{\downarrow}$ shown in
Fig. \ref{plot:FeCr_spin} indeed confirm the picture that evolved from
the BSF; i.e. $\sigma_{\downarrow}$ is about two orders of magnitude
larger than $\sigma_{\uparrow}$ and is nearly concentration
independent for $x$ $>$ 8\%. As a consequence the resistivity is
dominated by the majority spin channel.

In summary the resistivity increase from 0-10\% Cr (0-6\% V) is due
to the increased disorder scattering for the majority spin subsystem;
roughly speaking, the contribution of the majority subsystem to the
conductivity drops down. This drop down can be explained by a smeared
out BSF for this component. At higher Cr/V concentrations only the
minority subsystem contributes to the conductivity. The increase of
the Cr/V concentration leads to no broadening of the minority
states. Therefore, the contribution of this component to the
conductivity in the range of 10-20\% Cr (6-20\% V) is constant. This
leads to a nearly constant resistivity in that concentration regime.
\subsection{NL-CPA Results}
\begin{figure}
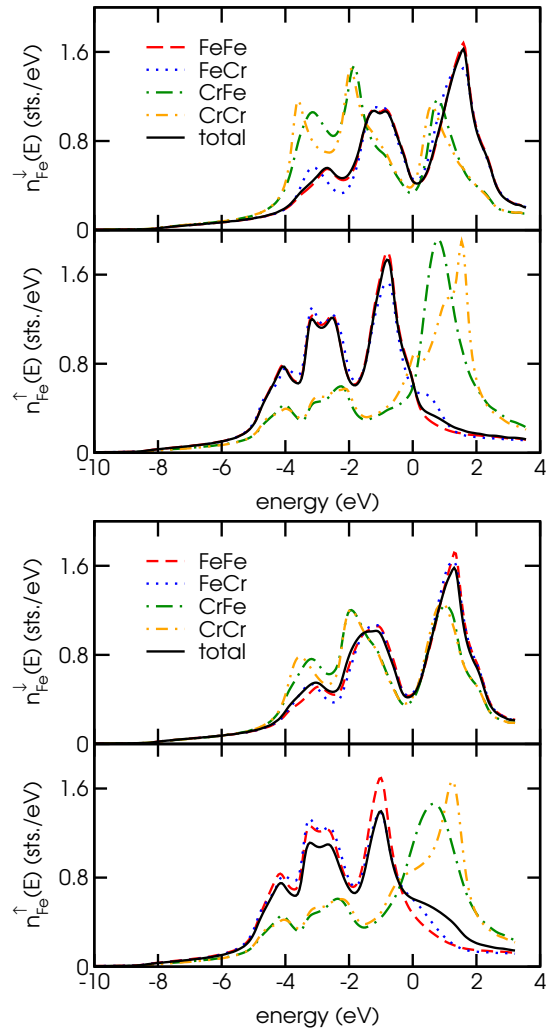

\begin{minipage}[b]{8.7cm}
 \includegraphics[scale=0.4,clip]{Fig10top_complete_scaled_same_P.eps}
 \includegraphics[scale=0.4,clip]{Fig10bottom_complete_0.2_scaled_nlcpa_pap_same_P.eps}
 \caption{\label{plot:DOS_NLCPA}(Color online) Cluster resolved NL-CPA density of states of disordered Fe$_{1-x}$Cr$_x$ (top: 4\% Cr, bottom: 20\% Cr). The solid (black) line shows the total DOS, the dashed (red) line shows the contribution of the FeFe-cluster, the (blue) dotted line of the FeCr-cluster, the (green) dashed-dotted line of the CrFe-cluster and the (orange) dashed-dotted-dotted line of the CrCr-cluster.}
\end{minipage}
\end{figure}

The next step in our analysis is to investigate the influence of SRO
effects on the residual resistivity as this was suggested to be the
reason for the anomalous concentration dependence of the residual
resistivity. To include SRO effects we use the NL-CPA which is, as
explained above, a cluster generalization of the CPA. Therefore we
have to define appropriate cluster configurations and their associated
probabilities ($P_{\gamma}$). We used the smallest possible bcc
cluster with two atoms. This leads to four different cluster
configurations e.g. Fe$_{1-x}$Cr$_x$: pure Fe (FeFe), pure Cr (CrCr)
and two mixed clusters with different occupation of the lattice sites
(FeCr, CrFe). The probabilities of the configurations depend on the
investigated ordering case. For example in the case of
Fe$_{0.5}$Cr$_{0.5}$ and SRO only the configurations FeCr and CrFe
have to be considered. For the simulation of clustering effects only
the pure configurations FeFe and CrCr have non-zero probabilities and
to simulate total disorder all configurations are used. In
Ref. \onlinecite{TSE+07} a detailed description how to define the
configuration probabilities for a bcc lattice is given. To display the
contribution of the different cluster configurations to the density of
states of a disordered Fe$_{1-x}$Cr$_x$ crystal we show in
Fig. \ref{plot:DOS_NLCPA} the cluster resolved density of states for
two different Cr concentrations. From this figure one can see that the
total DOS agrees very well with the total CPA DOS from
Figs. \ref{plot:DOS1} and \ref{plot:DOS}. The most dominant
contribution to the total DOS comes from the FeFe-cluster due to the
high Fe concentration. If one compares Fig. \ref{plot:DOS_NLCPA} with
Figs. \ref{plot:DOS1} and \ref{plot:DOS} one obtains a similar
behavior of the minority/majority spin subsystem. The minority part of
the DOS shows a hybridization between all cluster configurations
whereas the majority part shows a separation in energy between the
FeFe/FeCr- and CrCr/CrFe-clusters. \\
\begin{figure}
 \begin{center}
 \includegraphics[scale=0.36]{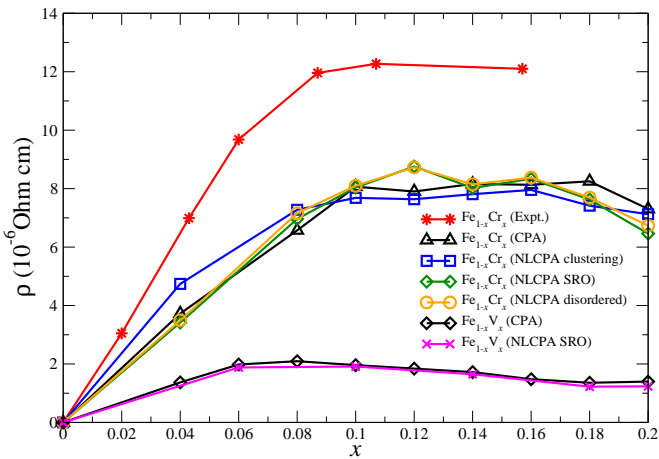}
 \caption{\label{plot:FeCr_res_nlcpa}(Color online) The resistivity of Fe$_{1-x}$Cr$_x$ and  Fe$_{1-x}$V$_x$ as function of the concentration $x$. Fe$_{1-x}$Cr$_x$: The asterisks (red line) show the experimental data of Ref. \onlinecite{Nikolaev99} and the triangles (black line) show our CPA results. The circles (yellow line) show the NL-CPA results for the disordered alloys, the diamonds (green line) show results when SRO is included and the squares (blue line) when short-range clustering is included.  Fe$_{1-x}$V$_x$: The diamonds (black line) show our CPA results and the crosses (magenta line) show results when SRO is included.}
 \end{center}
\end{figure}
Fig. \ref{plot:FeCr_res_nlcpa} shows our results for the residual
resistivity within the NL-CPA. In this plot we show again the three
curves from Fig. \ref{plot:FeCr_res} and additional some curves for
different ordering situations. To demonstrate the consistency of the
NL-CPA Kubo-Greenwood formalism we show in
Fig. \ref{plot:FeCr_res_nlcpa} the residual resistivity of randomly
disordered Fe$_{1-x}$Cr$_x$. These results agree for most
concentrations very well with the CPA results. This is satisfactory
because the CPA is a well established theory for the description of
disordered systems. \\ Experimentally it is observed that
Fe$_{1-x}$Cr$_x$ tends for $x < 0.1$ to SRO and for $x > 0.1$ to
clustering \cite{Mirebeau84} whereas Fe$_{1-x}$V$_x$ tends to SRO
\cite{TRS94}. Therefore we show in Fig. \ref{plot:FeCr_res_nlcpa}
NL-CPA calculations which simulate SRO (Fe$_{1-x}$Cr$_x$ and
Fe$_{1-x}$V$_x$) as well as clustering (Fe$_{1-x}$Cr$_x$). \\ The
important observation from our calculations is that the influence of
ordering effects (SRO and clustering) on the residual resistivity is
very small for these systems. In Ref. \onlinecite{TSE+07} the same
formalism was applied to bcc CuZn and gave a strong variation of the
resistivity as function of the ordering state. The small influence of
ordering effects on the resistivity for Fe$_{1-x}$Cr$_x$ was also
found experimentally by \citet{Mirebeau84}. This shows that the
formalism from Ref. \onlinecite{TSE+07} can handle various different
ordering dependences of the residual resistivity.
\section{Discussion and Summary}
The present work deals on an $ab$-$inito$ level with the anomalous
concentration dependence of the residual resistivity in
Fe$_{1-x}$Cr$_x$ and Fe$_{1-x}$V$_x$ alloys ($x$ $\le$ 0.2). Within
CPA we obtain a Nordheim like behavior for small Cr/V concentrations
(smaller $\approx 8\%$ Cr, smaller $\approx 6\%$ V) and an
approximately constant residual resistivity of $8\,\mu \Omega$cm in
the regime from 10-20\% Cr and $2\,\mu \Omega$cm in the regime from
6-20\% V, respectively. Such a concentration dependence of the
residual resistivity can be explained by the different contributions
of the majority/minority spin channel to the conductivity. The BSF of
these spin channels are affected differently by the increase of the
Cr/V concentration as shown in Fig. \ref{plot:BSF} and
Fig. \ref{plot:BSF_FeV}. The broadening of the majority channel leads
to a lifetime of these states approaching zero. Therefore, the
contribution of the majority channel to the conductivity drops down,
whereas the BSF for the minority channel is nearly unaffected by the
increase in Cr/V concentration. Hence, the contribution of these
states to the conductivity is constant and responsible for the plateau
in the regime from 10-20\% Cr (6-20\% V). These observations explain
the anomalous resistivity behavior of Fe-rich antiferromagnetic alloys
with comparable electronic structure properties as Fe$_{1-x}$Cr$_x$
and Fe$_{1-x}$V$_x$, respectively. It turns out, that the
concentration independence of the residual resistivity is by no means
anomalous for these materials. Therefore, we predict a similar
behavior for other alloy systems from the same branch of the
Slater-Pauling plot.\\ In addition, we investigated the influence of
short-ranged correlations in the lattice site occupation by employing
the NL-CPA formulation of the Kubo-Greenwood equation. The inclusion
of such short-range effects has only little influence on the
results.\\ The comparison of the Fe$_{1-x}$Cr$_x$ calculations with
experiment show satisfying agreement. The difference in the height of
the plateau, as compared to the experiment, could be attributed to
impurities, lattice defects, grain boundaries, etc. which are always
present in samples and therefore contained in the experimental
data. Such imperfections, which have been neglected in the present
calculations, lead in general to an increase of the measured
resistivity \cite{Ros87}. \\ Our calculations show the initial linear
increase albeit with a lower slope than seen in experiment. In
Ref. \onlinecite{Tsiovkin05} it is argued that this is a consequence
of the limitations of the CPA for alloys with a dominant concentration
of one constituent. However, it should be pointed out that our NL-CPA
results confirm the single-site CPA data.\\ Our calculations reveal a
plateau of the residual resistivity starting at the same Cr
concentration as seen in experiment (at 10\% Cr). This is in variance
to an earlier theoretical study \cite{Tsiovkin05} which finds the
starting point of the plateau only at $\approx$ 20\% Cr.
\section{Acknowledgments}
S.L. would like to thank the SFB 689 ``Spinph\"anomene in
reduzierten Dimensionen'' for financial support. D.K. acknowledges
support from the priority program of the DFG  SPP 1145
 ``Modern and universal first-principles methods for
many-electron systems in chemistry and physics''.

\newpage 

\end{document}